# Prediction of Nontrivial Band Topology and Superconductivity in Mg$_2$Pb

Guang Bian,[1*] Tay-Rong Chang,[2*] Angus Huang,[3,4*] Yuwei Li,[1] Horng-Tay Jeng,[3,4] David J. Singh,[1] Robert J. Cava,[5] and Weiwei Xie [6]

[1.] Department of Physics and Astronomy, University of Missouri, Columbia, Missouri 65211, USA
[2.] Department of Physics, National Cheng Kung University, Tainan, 701, Taiwan
[3.] Department of Physics, National Tsing Hua University, Hsinchu 30013, Taiwan
[4.] Institute of Physics, Academia Sinica, Taipei 11529, Taiwan
[5.] Department of Chemistry, Princeton University, Princeton, New Jersey 08544, USA
[6.] Department of Chemistry, Louisiana State University, Baton Rouge, Louisiana 70803, USA

**Abstract**

**The interplay of BCS superconductivity and nontrivial band topology is expected to give rise to opportunities for creating topological superconductors, achieved through pairing spin-filtered boundary modes via superconducting proximity effects. The thus-engineered topological superconductivity can, for example, facilitate the search for Majorana fermion quasiparticles in condensed matter systems. Here we report a first-principles study of Mg$_2$Pb and predict that it should be a superconducting topological material. The band topology of Mg$_2$Pb is identical to that of the archetypal quantum spin Hall insulator HgTe, while isostructural and isoelectronic Mg$_2$Sn is topologically trivial; a trivial to topological transition is predicted for Mg$_2$Sn$_{1-x}$Pb$_x$ for x~0.77. We propose that Mg$_2$Pb-Mg$_2$Sn quantum wells should generate robust spin-filtered edge currents in analogy to HgTe/CdTe quantum wells. In addition, our calculations predict that Mg$_2$Pb should become superconducting upon electron doping. Therefore, Mg$_2$Pb is expected to provide a practical material platform for studying emergent phenomena arising from the interplay of superconductivity and band topology.**

*these authors equally contributed to this work.

## Introduction

Recently, significant research effort had been devoted to the search for materials that incorporate both superconductivity and the topological band structure that is found in topological insulators (TIs)[1-4]. The interplay of nontrivial band topology and superconductivity is expected to give rise to circumstances where time-reversal-invariant $p \pm ip$ superconductivity can naturally emerge as a consequence of the spin-momentum locking of topological surface electrons[5]. The surface $p \pm ip$ superconductivity in such superconductors would create a solid-state environment for realizing the Kitaev picture of Majorana fermions[6]. Several materials hosting both nontrivial band topology and superconductivity have been discovered and synthesized experimentally[7-20]. Generally, these materials belong to two categories. The first are artificially fabricated TI/superconductor heterostructures in which superconductivity is introduced into TI films through superconducting proximity effects[5,18,20]. The others are doped topological insulators that become superconducting themselves below a critical temperature[8,9,13,16,21]. Although superconductivity has been found in these topological materials, new systems can be expected to display different characteristics. Therefore, there is a need for the identification of new topological superconductors. In this paper, we show by using first-principles calculations that dimagnesium plumbide, $Mg_2Pb$, is expected to be a BCS-type superconductor on electron doping, and, at same time, possesses a simple topological band structure that resembles that of the quantum spin Hall parent compound HgTe[22]. Further a racial to topological transition is predicted for the $Mg_2Pb_{1-x}Sn_x$ system. Our theoretical results indicate that $Mg_2Pb$ provides a promising alternative material path to search for topological superconductivity and Majorana quasiparticles in condensed matter systems.

## Results and Discussion

$Mg_2Pb$ crystallizes in a cubic anti-fluoride structure in which Pb atoms form a face-centered cubic (FCC) arrangement and the magnesium atoms occupy the eight tetrahedral voids, as shown in Fig. 1(*a*). The cubic lattice constants (in space group Fm3m, # 225) are 6.815 Å and 6.776 Å for $Mg_2Pb$ and the isostructural compound $Mg_2Sn$, respectively[23-25]. Figure 1(*b*) shows the first Brillouin zone of a FCC lattice with the high symmetry *k* points marked. The calculated bulk band structure of $Mg_2Pb$, obtained here using the generalized gradient approximation (GGA) method with the inclusion of spin-orbit coupling (SOC) is shown in Fig. 1(*c*). The Fermi surface is predominantly comprised of a hole pocket around Γ and an electron pocket around L. The calculated density of states indicates that the Pb-6*p* and Mg-3*s* orbitals dominate the conduction and valence bands close to the Fermi energy $E_F$. The overall band dispersion of $Mg_2Pb$ is closely analogous to that of the archetypal quantum spin Hall compound mercury telluride (HgTe) except that the conduction band along Γ-L dives down to the Fermi level, forming an electron pocket at L. In contrast to the point-like Fermi surface of HgTe[26], the Fermi surface of $Mg_2Pb$ is large enough to allow for an effective Cooper pairing instability for low-energy electrons. Stoichiometric $Mg_2Pb$ was reported to be superconducting in an old compilation[27], but that report was in error.

We now focus on the symmetry and topology of the $Mg_2Pb$ band structure. Figure 2(*a*) shows the orbital decomposition of each energy band into Pb-6*p* and Mg-3*s* orbitals, which constitute the dominant low-energy electronic states. There are three sets of bands close to the Fermi level of Mg$_2$Pb. For convenience, we employ the representation notation of HgTe to label the three relevant bands of Mg$_2$Pb[26,28], because the two compounds essentially share the same band topology. As marked in Fig. 2(*a*), States $\Gamma_6$, $\Gamma_7$, and $\Gamma_8$ are two-, two-, and fourfold degenerate at $\Gamma$, and the $\Gamma_8$ state splits into two branches as it disperses away from $\Gamma$ due to reduced symmetry. Each branch is doubly degenerate due to the presence of time-reversal and space inversion symmetries. The $\Gamma_8$ states lie above the Fermi level at $\Gamma$. The hole-like branch of $\Gamma_8$ is primarily derived from Pb-6*p* orbitals while the electron-like branch evolves gradually from Pb-6*p* to Mg-3*s* with ***k*** moving away from $\Gamma$. The $\Gamma_6$ band grazes the Fermi level at $\Gamma$. In contrast to $\Gamma_8$, the $\Gamma_6$ state at $\Gamma$ originates from the Mg-3*s* orbital. This leads to a discrepancy in parity eigenvalue between $\Gamma_6$ and $\Gamma_8$ states. Our calculation shows that the parity eigenvalues are even and odd for $\Gamma_6$ and $\Gamma_8$ states, respectively. It is this ordering of bands with opposite parity eigenvalues that gives rise to a nontrivial band topology, analogous to what happens in HgTe. The hole-like $\Gamma_7$ band, mainly from Pb-6*p* orbitals, is 1 eV below the Fermi energy and is irrelevant to the band topology near the Fermi energy.

The four-fold degeneracy of $\Gamma_8$ is protected by the $C_4$ symmetry of the lattice and therefore can be only lifted by a symmetry-breaking perturbation. In order to simulate a symmetry-breaking effect, we performed calculations where we compressed the cubic lattice by 3% along the cubic body diagonal direction, which is equivalent to increasing the angle *α* between FCC primitive lattice vectors from 60º to 61º. The resulting band structure is plotted in Fig. 2(*b*). Indeed, the $C_4$-symmetry-breaking strain induces a gap of size 0.1 eV at the degenerate point of the $\Gamma_8$ band. With this gap, the conduction band is separated from the valence band throughout the whole Brillouin zone and, therefore, the topological $Z_2$ invariant can be calculated for this centrosymmetric compound by examining the parity eigenvalues of the valence band at time-reversal invariant momenta (TRIM) points[29]. We find that the $Z_2$ invariant is $v = -1$, identical to that of strained HgTe and topological insulators, indicating that cubic Mg$_2$Pb has a topologically nontrivial band structure.

The defining characteristic of a topological insulator is the existence of topological surface states (TSS). To explore the surface states for Mg$_2$Pb, we calculated the band structure of a 30-unit cell undistorted cubic Mg$_2$Pb slab with a (110) surface. The result is shown in Fig. 2(c). The bands are plotted with the weight of the wavefunction on the top layer of the slab indicated in color, which makes it easy to discern the surface vs. bulk character of a state. Despite the absence of a bulk band gap at $\bar{\Gamma}$, a surface band is visible within a partial bulk band gap around the $\bar{Y}$ point of the (110)-surface Brillouin zone. The surface band possesses a large Rashba spin splitting as a result of the broken space inversion symmetry at the surface and the strong atomic spin-orbit coupling (SOC) of the Pb atoms. We find that $E_R$ = 29.5 meV, $k_o$ = 0.063 Å$^{-1}$ and $\alpha_R = 2E_R/k_o$ = 0.93 eV·Å, which is comparable with some of the known bulk Rashba semiconductors such as BiTeI ($E_R$ = 100 meV, $\alpha_R$ = 3.8 eV·Å) and BiTeCl ($E_R$ = 18.5 meV, $\alpha_R$ = 1.2 eV·Å)[30-32]. The Rashba point appears 0.4 eV below the Fermi level at $\bar{Y}$ and the tails of the surface band merge into the bulk bands at $\bar{\Gamma}$. The vanishing bulk band gap in the undistorted material obscures a clear identification of the band topology. Lattice strains can exist in epitaxial films grown on substrates, however, which would lead to a desirable band gap at the degenerate point of the $\Gamma_8$ band[26,33,34]. To simulate the strain effect, we calculated the electronic band structure of a semi-infinite Mg$_2$Pb-(110) slab with a 2% lattice compression along the (1-10)-direction, which corresponds to $\bar{\Gamma} - \bar{X}$ in $k$ space. The band structure along $\bar{\Gamma} - \bar{Y}$ colored with the surface weight of each state (Figs. 3(a, b)) shows that a band gap of about 30 meV is induced by such a strain, and that a Dirac surface band emerges inside the band gap in the same way as is the case for strained HgTe and $\alpha$-Sn films[33-37]. The Dirac surface band traverses the band gap in the strained material in a gapless manner due to its topological origin. The topological character is further corroborated by inspecting the calculated spin texture of the surface band, which is shown in Figs. 3(c, d). The calculated dominant spin component of the surface states is $S_X$, which is in-plane and parallel to $\bar{\Gamma} - \bar{X}$. The other two spin components are found to be negligible. Therefore, the calculated spin polarizations of the Dirac surface states are perpendicular to the direction of the momentum, thus exhibiting a spin-momentum locking configuration which is characteristic of topological surface states.

Another way to visualize the topological nature of the $Mg_2Pb$ band structure is to vary the effective SOC, a critical factor in this compound. Thus in Figures 4(*a*) and 4(*b*) we show the calculated band structures of $Mg_2Pb$ without and with the inclusion of SOC. In the absence of SOC, the $\Gamma_6$ band lies above the $\Gamma_8$ band. The $\Gamma_7$ and $\Gamma_8$ bands are degenerate at $\Gamma$. When SOC is turned on, $\Gamma_6$ and $\Gamma_8$ are inverted in energy at $\Gamma$. The nontrivial band topology of $Mg_2Pb$ arises from this band inversion. At the same time, $\Gamma_7$ is split off from $\Gamma_8$ by SOC, pushing $\Gamma_7$ to a higher energy region. The calculated bands for $Mg_2Sn$ obtained similarly are presented in Figs. 4(*c*) and 4(*d*) for comparison. Since Sn is lighter than Pb, the atomic SOC of $Mg_2Sn$ is weaker than that of $Mg_2Pb$. The band ordering of $Mg_2Sn$ at the zone center is same as in the case of $Mg_2Pb$ without SOC, that is $\Gamma_6$ is higher in energy than $\Gamma_8$. Therefore, the energy difference $\Delta E = E_6(\Gamma)-E_8(\Gamma)$ is positive for $Mg_2Sn$ but negative for $Mg_2Pb$. This change in sign signifies that $Mg_2Sn$ belongs to a topologically distinctive topological phase. In other words, the band inversion and nontrivial topology of $Mg_2Pb$ is induced by the strong SOC of the Pb atoms. $Mg_2Sn$ and $Mg_2Pb$ form an isostructural chemical pair with different band ordering, similar to the quantum spin Hall parent compounds HgTe and CdTe. The effective SOC can be tuned by varying the chemical composition in $Mg_2Sn_{1-x}Pb_x$. At the critical composition $x_c$, the band gap vanishes, and a topological phase transition occurs. Figure 4(e) presents the bulk energy gap as a function of Pb concentration $x$ calculated via a mixed pseudopotential method[38]. The critical concentration for the trivial to topological transition is found to be $x_c = 0.77$. The band structure at the critical point, presented in the upper panel of Fig. 4(e), shows the "touching" of $\Gamma_6$ and $\Gamma_8$ bands, and a linear bulk band dispersion, which is different from the symmetry-protected three-band crossings.[39] The system becomes a 3D Dirac semimetal at the critical point[40]. The chemical and structural affinity between $Mg_2Pb$ and $Mg_2Sn$, suggests that a quantum well heterostructure consisting of $Mg_2Sn/Mg_2Pb/Mg_2Sn$ with a variable well width, should achieve the quantum spin Hall effect with a robust spin-momentum-locked edge current, analogous to what has been done for CdTe/HgTe/CdTe quantum wells[22]. What makes $Mg_2Pb$ different from HgTe is that with electron doping $Mg_2Pb$ may become a BCS-type superconductor.

In order to explore the possible existence of superconductivity in $Mg_2Pb$, we performed systematic calculations for the electron-phonon coupling. Figure 5(a) shows the calculated phonon spectrum of undoped cubic $Mg_2Pb$, in which the magnitude of the calculated electron-phonon coupling $\lambda_{qv}$ is indicated by the size of red dots. In our calculations, we find the total electron-phonon coupling $\lambda = \sum_{qv} \lambda_{qv}$ to be 0.239. The major contribution to $\lambda$ comes from the acoustic phonon mode with lowest energy. Using the McMillan formula (see Method section for details), the superconducting transition temperature $T_c$ is estimated to be 0.002K for pure $Mg_2Pb$, an extremely low temperature. $Mg_2Sn$ is not calculated to be superconducting at all. Our calculations show, however, that λ and $T_c$ increase significantly when doping $Mg_2Pb$ with electrons (or, equivalently, raising the Fermi level). The calculated relation between the electron doping and the Fermi level shift is illustrated in Fig. 5(b), where negative values of electron doping correspond to hole doping. We calculated λ and $T_c$ at several doping levels as marked in the band structure shown in Fig. 5(c). The results in Fig. 5(d) suggest a monotonic increase of the superconducting critical temperature on raising the Fermi level. For example, when adding 0.7 electrons to the unit cell, $T_c$ is calculated to become approximately 1.4 K. This effect becomes more prominent when the Fermi level shift is larger than 0.4 eV. Even in the event that the superconducting critical temperature is not accurately estimated in the calculations, the relative increase of $T_c$ by a factor of 1.4/0.002 ~ 700 on electron doping of $Mg_2Pb$ may bring any superconducting transition in this system into an experimentally observable temperature range. On electron doping, the Fermi level is close to the band edge of the two $\Gamma_8$ branches, so the enhancement of superconductivity is likely due to the participation of electrons from the $\Gamma_8$ band edge and the electron-like $\Gamma_8$ band in the superconducting Cooper pairing. Electron doping also should also make possible the placing of the Femi level inside the topological band gap of a strained $Mg_2Pb$ film, in turn pairing the topological surface electrons through an effective surface-bulk proximity effect[41]. Therefore, this coincidence between the Fermi level shift and the strain-induced topological band gap is predicted to give rise to a potential visualization of helical $p \pm ip$ superconductivity on the surface of $Mg_2Pb$[18]. Hole doping, by contrast, cannot promote the superconductivity in the compound.

**Conclusion**

In summary, we have investigated electronic structure and electron-phonon coupling of $Mg_2Pb$ and $Mg_2Sn$ by first-principles calculations. Our calculations reveal that $Mg_2Pb$ should be a BCS-type superconductor upon electron doping, and at the same time possesses a nontrivial electronic band topology thanks to the strong spin-orbit coupling of Pb atoms. On the other hand, $Mg_2Sn$ is calculated to be topologically trivial with a band ordering identical to that of CdTe. The calculated coexistence of nontrivial band topology and superconductivity makes $Mg_2Pb$ a promising material for hosting topological superconductivity. There are two ways to visualize the topological superconductivity in $Mg_2Pb$. The first way is to create an $Mg_2Sn/Mg_2Pb/Mg_2Sn$ quantum well heterostructure, as has been done for CdTe/HgTe/CdTe quantum wells. By controlling the width of the quantum well, one may be able to tune the system into the quantum spin Hall phase and thus generate a pair of spin-momentum-locked topological edge bands. Because $Mg_2Pb$ is calculated to be superconducting on electron doping, the edge state electrons may pair up through superconducting proximity effects to create a 1D $p \pm ip$ superconductor. The other approach would be to shift the Fermi level to the energy of the Dirac surface states in a strained $Mg_2Pb$ film by substitutional alloying or applying a gating voltage. The spin-polarized surface electrons may then form Cooper pairs facilitated by the superconducting bulk-surface proximity effect, and, thus, generate a 2D topological $p \pm ip$ superconductor, which is line with the Fu-Kane mechanism of proximity-effect induced p-wave superconductivity[5]. Taking collectively our results on calculated band topology and superconductivity, we propose that $Mg_2Pb$ provides a versatile platform for a material realization of topological superconductivity, especially in low dimensions. In addition, we calculate that by tuning the composition in $Mg_2Sn_{1-}$

$_x$Pb$_x$ to the critical value, one can potentially obtain a superconducting Dirac semimetal with a linear band dispersion, therefore opening a door for studying the rich physics arising from the interplay between superconductivity and Dirac semimetallic states.

**Methods**

We computed electronic structures using the projector augmented wave method [42,43] as implemented in the VASP package[44-46] within the generalized gradient approximation (GGA) [47] schemes. Experimental lattice constants were used and Monkhorst-Pack *k*-point meshes were used in the bulk and slab calculations, respectively. The spin-orbit coupling effects are included self-consistently.

The electron-phonon coupling, $\lambda_{qv}$ is computed based on density functional perturbation theory[48] implemented in the Quantum Espresso code [49], using

$$\lambda_{qv} = \frac{1}{\pi N_F} \frac{\Pi''_{qv}}{\omega^2_{qv}} \qquad (1)$$

Where $N_F$ is the density of states (DOS) at the Fermi level, and $\lambda_{qv}$ is the phonon frequency of mode *v* at wave vector *q*. The electron-phonon quasiparticle linewidth, $\Pi''_{qv}$, is given by

$$\Pi''_{qv} = \pi \omega_{qv} \sum_{mn,k} |g^v_{mn}(\mathbf{k},\mathbf{q})|^2 \, \delta(\epsilon_{nk}) \delta(\epsilon_{mk+q}) \qquad (2)$$

Where $\epsilon_{nk}$ is the energy of the KS orbital and the dynamical matrix reads:

$$g^v_{mn}(\mathbf{k},\mathbf{q}) = \left(\frac{\hbar}{2M\omega_{qv}}\right)^{1/2} \left\langle \psi_{nk} \left| \frac{dV_{scf}}{du_{qv}} \cdot \hat{e}_{qv} \right| \psi_{mk+q} \right\rangle \qquad (3)$$

Where $\frac{dV_{scf}}{du_{qv}}$ represents the deformation potential at the small atomic displacement $d\mathbf{u}_{qv}$ of the given phonon mode. M and $\hat{e}_{qv}$ denote the mass of the atom and the unit vector alone $\mathbf{u}_{qv}$, respectively. The critical temperature $T_c$ can then be estimated by McMillan formula:

$$T_c = \frac{\omega_{ln}}{1.20} \exp\left[-\frac{1.04(1+\lambda)}{\lambda - \mu^*(1+0.62\lambda)}\right], \qquad (4)$$

where

$$\lambda = \sum_{qv} \lambda_{qv} \qquad (5)$$

$$\omega_{ln} = \exp\left[\frac{2}{\lambda} \int d\omega \frac{\ln \omega}{\omega} \alpha^2 F(\omega)\right] \qquad (6)$$

$$\alpha^2 F(\omega) = \frac{1}{2} \int_{BZ} d\omega \lambda_{qv} \omega_{qv} \delta(\omega - \omega_{qv}) \qquad (7)$$

$$\mu^* = 0.1 \qquad (8)$$

**Acknowledgements**

GB acknowledges financial support from University of Missouri, Columbia. DJS and YL were supported by DOE through the computational materials science MAGICS center, DE-SC0014607. The work at Princeton University was supported by the ARO MURI on topological insulators, grant W911NF-12-1-0461. WX was supported by the LSU-startup funding. TRC was supported by the Ministry of Science and Technology and National Cheng Kung University, Taiwan. HTJ was supported by the Ministry of Science and Technology, National Tsing Hua University, and Academia Sinica, Taiwan. TRC and HTJ also thank NCHC, CINC-NTU, and NCTS, Taiwan for technical support.

**Author Contribution**

WX and GB searched for the materials candidates and designed the research; GB, TRC, AH, WX, HTJ YL and DJS conducted the computational work and RJC contributed to the analysis of the results. WX supervised the research. All authors contributed to the paper writing.

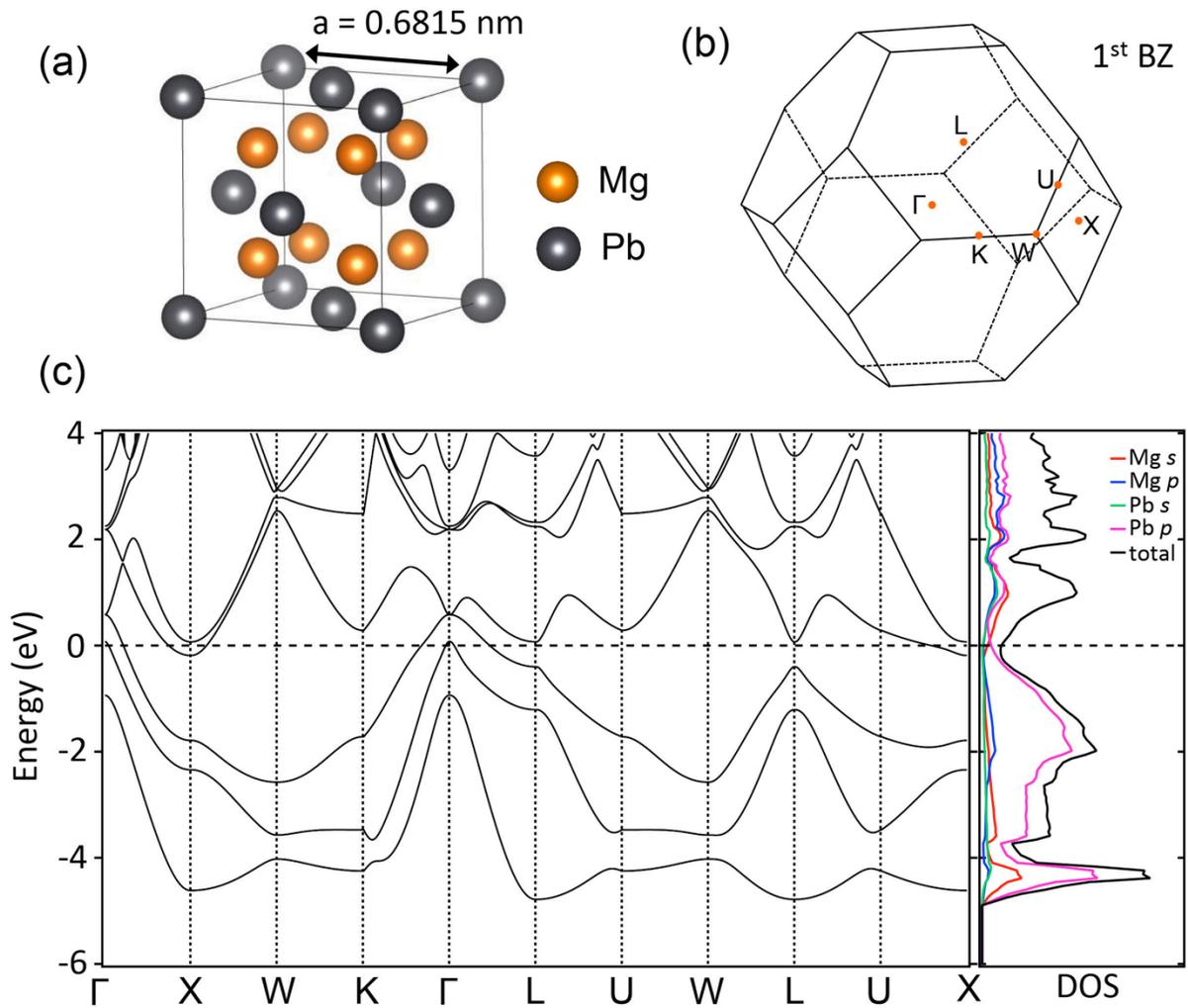

**Figure 1.** (*a*) The cubic anti-fluorite lattice structure of undistorted $Mg_2Pb$. (*b*) The first Brillouin zone with high symmetry points indicated. (c) Bulk band structure and density of states (DOS) of cubic $Mg_2Pb$ using the method of generalized gradient approximation (GGA). The spin-orbit coupling (SOC) is taken into account in the calculation. The orbital-projected DOS curves are plotted in color.

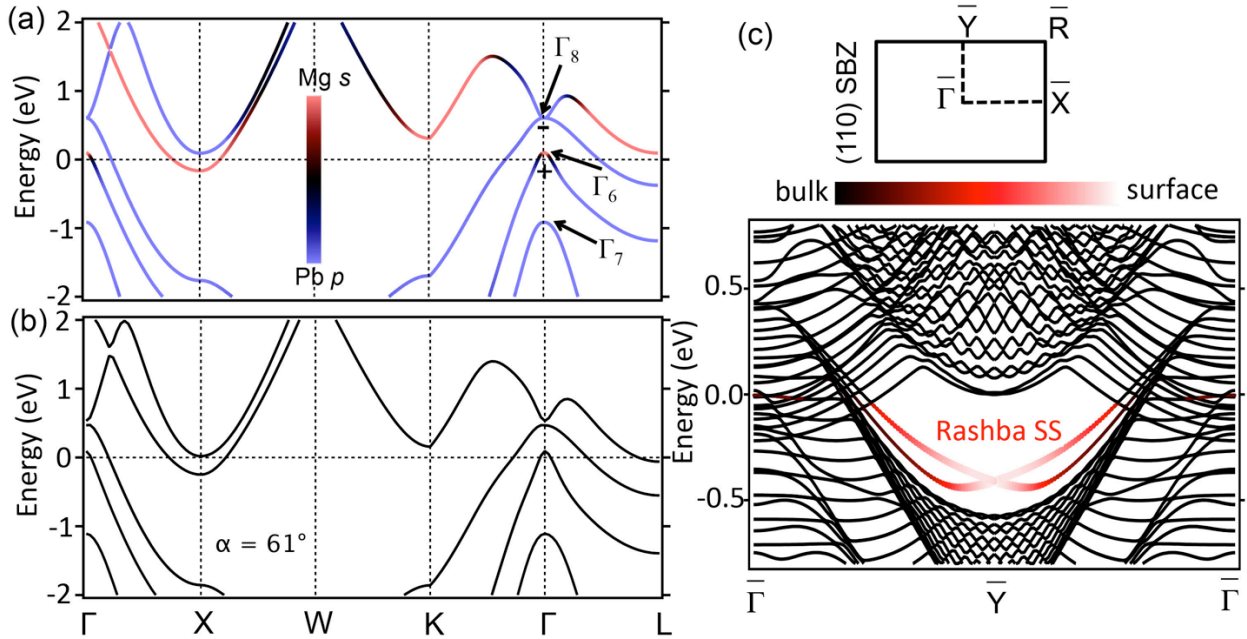

**Figure 2.** (*a*) Band structure of cubic undistorted $Mg_2Pb$ with orbital projection onto Mg-s and Pb-p orbitals. The color bar gives the projection weight. The symmetry of each band is denoted by $\Gamma_i$ ($i$ = 6; 7; 8). The '+' and '-' signs indicate the parity eigenvalues of corresponding states. (*b*) The bulk band structure of distorted $Mg_2Pb$ with a strain compressing the cubic lattice along the body diagonal direction by 3 percent. The angle between FCC primitive vectors is increased from 60° to 61° by the strain. (*c*) Band structure of an undistorted 30-unit-cell $Mg_2Pb$-(110) slab and the surface Brillouin zone. $\bar{\Gamma} - \bar{X}$ and $\bar{\Gamma} - \bar{Y}$ correspond to the (1-10) and (001) directions, respectively. The bands are colored according to the weight of wavefunction on the top layer of the slab.

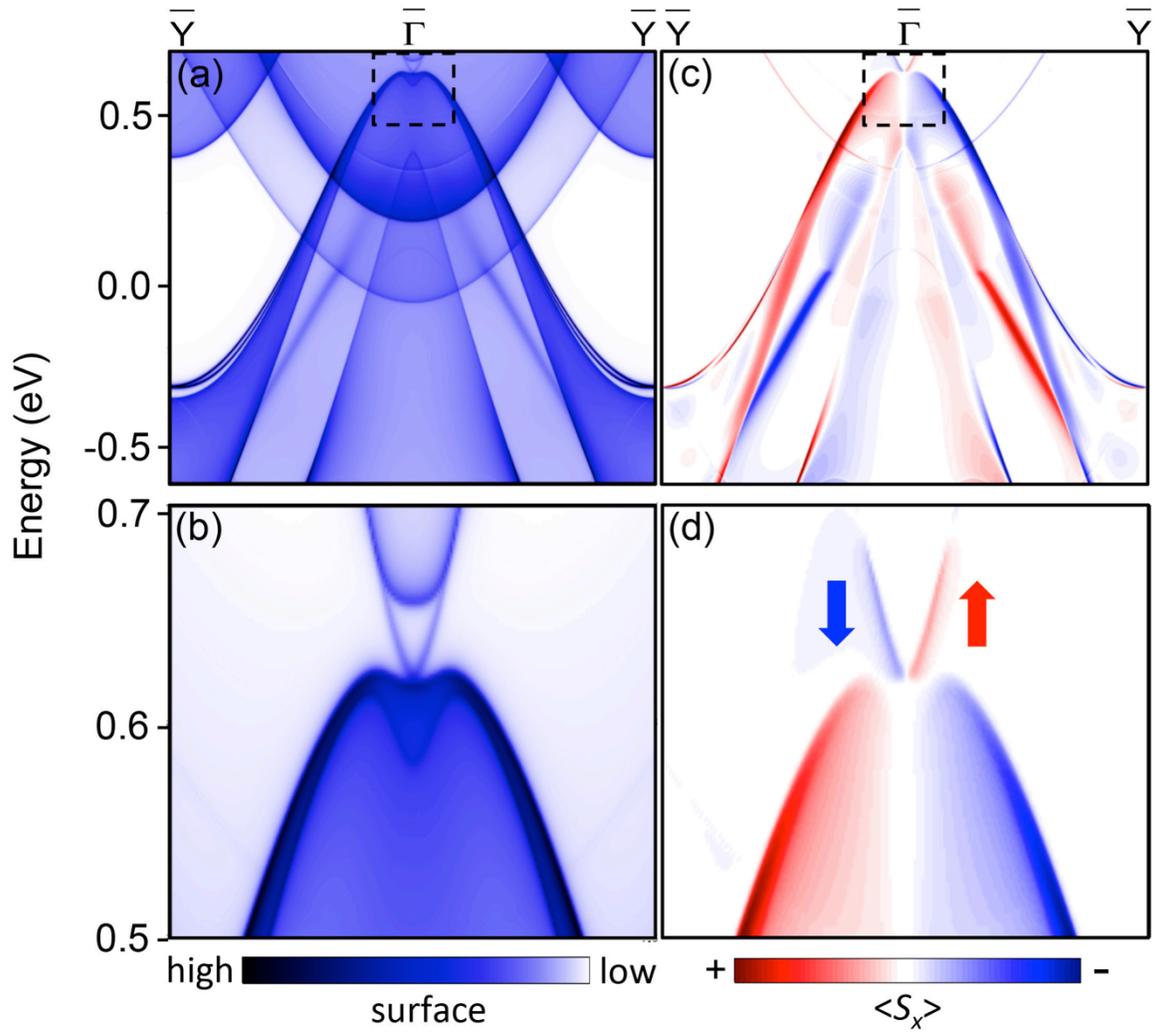

**Figure 3.** (*a*) Calculated band structure of a semi-infinite $Mg_2Pb$-(110) slab with a 2% lattice compression along the (1-10) direction. The depth of blue color shows the surface weight of each wave function. (*b*) Zoom-in of the part of the electronic band structure enclosed by the dashed rectangle in (a). (*c* and *d*) Spin polarization of bands shown in red and blue. The spin is aligned primarily along $S_X$, i.e., the spin is in-plane and perpendicular to the direction of momentum. The other two components of spin are negligibly small.

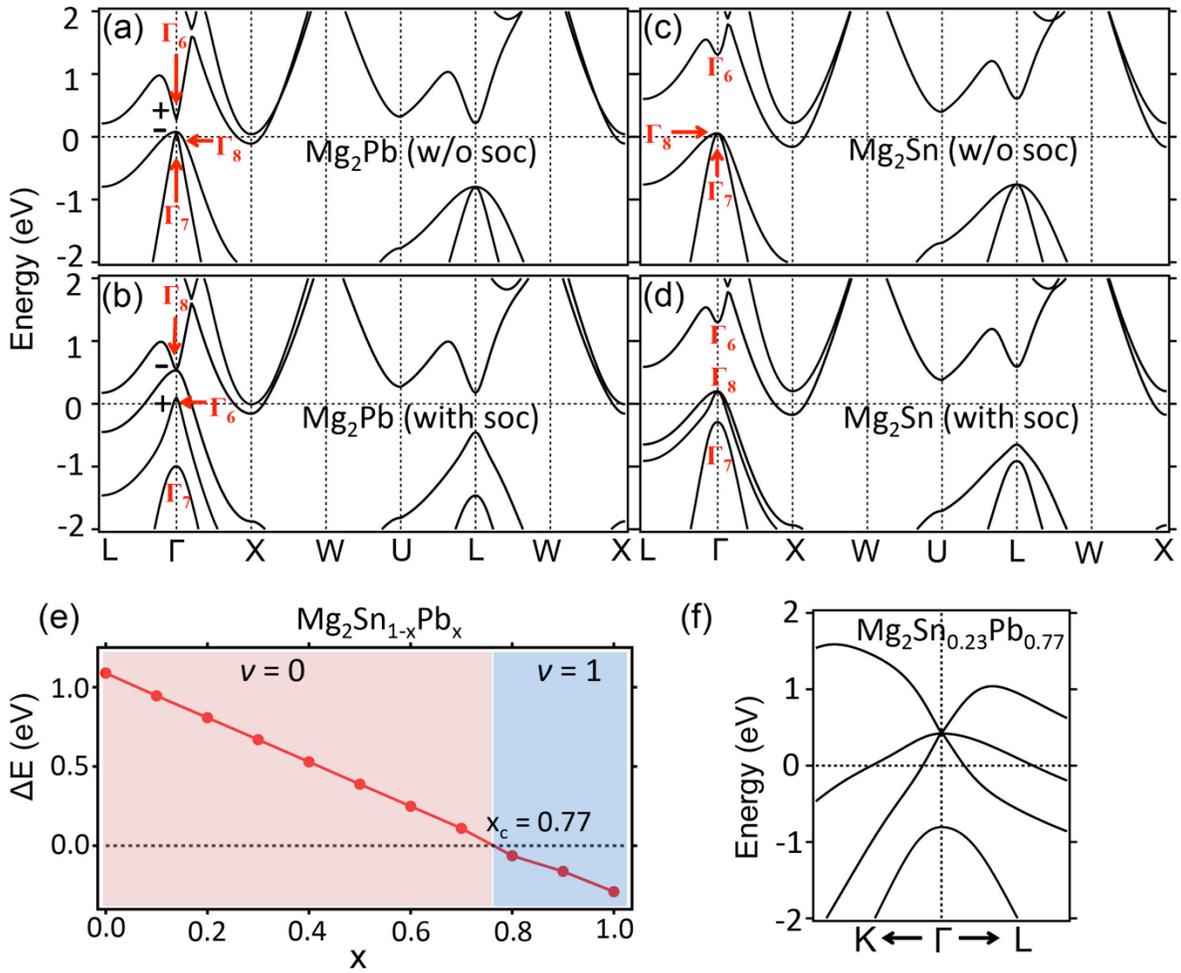

**Figure 4.** Calculated band structure of (*a*) undistorted cubic $Mg_2Pb$ without SOC, (*b*) same for $Mg_2Pb$ with SOC, (*c*) $Mg_2Sn$ without SOC, and (*d*) $Mg_2Sn$ with SOC. (*e*) The phase diagram of $Mg_2Sn_{1-x}Pb_x$ with the bulk band gap as a function of the composition $x$. The critical composition is $x_c = 0.77$. (*f*) The band structure of $Mg_2Sn_{1-x}Pb_x$ at $x_c = 0.77$.

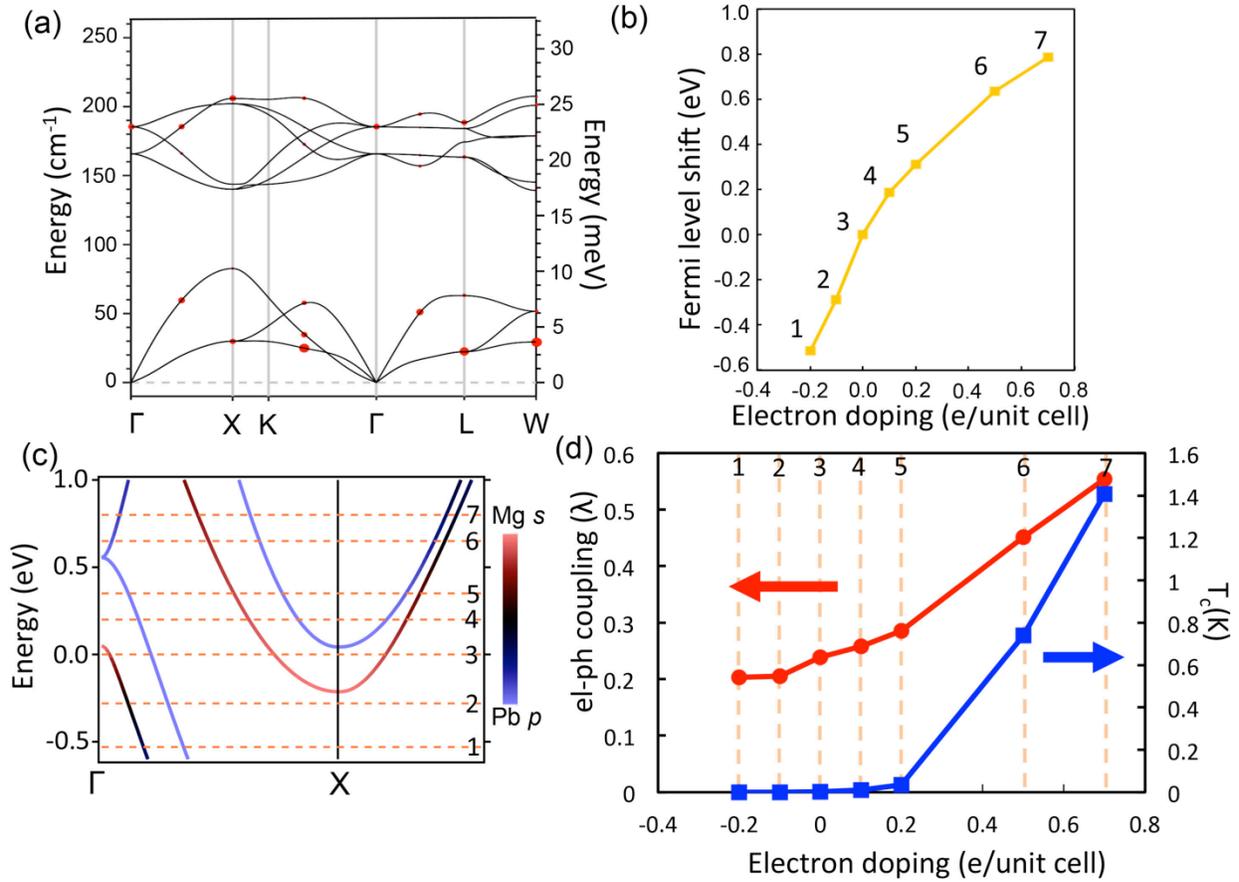

**Figure 5.** (*a*) calculated Phonon dispersion for cubic Mg$_2$Pb. The size of red dots indicates the magnitude of the electron-phonon coupling $\lambda_{qv}$ at different places in the Brillion Zone (*b*) The dependence of Fermi level shift on electron doping. (*c*) The band structure of Mg$_2$Pb. The energies marked by the dashed lines correspond to the results in (*b*, *d*) with same labels. (*d*) The calculated superconducting critical temperature T$_c$ and effective electron-phonon coupling $\lambda_{qv}$ of Mg$_2$Pb as a function of electron doping concentration.